\documentclass{article}
\usepackage{amssymb}
\usepackage{amsmath}
\usepackage{amsmath}
\usepackage{amsmath}
\usepackage{amsmath}
\usepackage{amsmath}
\usepackage{amsmath}
\usepackage{amsmath}
\usepackage{amsmath}
\usepackage{amsmath}
\usepackage{amsmath}
\usepackage{amsmath}
\usepackage{amsmath}
\usepackage{amsmath}
\usepackage{amsmath}
\usepackage{amsmath}
\usepackage{amsmath}
\usepackage{amsmath}
\usepackage{amsmath}
\usepackage{amsmath}
\usepackage{amsmath}
\usepackage{amsmath}
\usepackage{amsmath}
\usepackage{amsmath}
\usepackage{amsmath}
\usepackage{amsmath}
\usepackage{amsmath}

\input{tcilatex}

\begin{document}

\title{{\Large Path integral representations in noncommutative quantum
mechanics and noncommutative version of Berezin-Marinov action.}}
\author{D.M. Gitman$^{1}$, V.G. Kupriyanov$^{1,2}$ \\
\\
$^{1}$Instituto de F\'{\i}sica, Universidade de S\~{a}o Paulo, Brazil\\
$^{2}$ Physics Department, Tomsk State University,\textit{\ }Russia\\
E-mail: gitman@dfn.if.usp.br (D.M.Gitman), kvg@dfn.if.usp.br
(V.G.Kupriyanov).}
\date{\today                                        }
\maketitle

\begin{abstract}
It is known that actions of field theories on a noncommutative space-time
can be written as some modified (we call them $\theta $-modified) classical
actions already on the commutative space-time (introducing a star product).
Then the quantization of such modified actions reproduces both space-time
noncommutativity and usual quantum mechanical features of the corresponding
field theory. The $\theta $-modification for arbitrary finite-dimensional
nonrelativistic system was proposed by Deriglazov (2003). In the present
article, we discuss the problem of constructing $\theta $-modified actions
for relativistic QM. We construct such actions for relativistic spinless and
spinning particles. The key idea is to extract $\theta $-modified actions of
the relativistic particles from path integral representations of the
corresponding noncommtative field theory propagators. We consider
Klein-Gordon and Dirac equations for the causal propagators in such
theories. Then we construct for the propagators path-integral
representations. Effective actions in such representations we treat as $%
\theta $-modified actions of the relativistic particles. To confirm the
interpretation, we quantize canonically these actions. Thus, we obtain the
Klein-Gordon and Dirac equations in the noncommutative field theories. The $%
\theta $-modified action of the relativistic spinning particle is just a
generalization of the Berezin-Marinov pseudoclassical action for the
noncommutative case.
\end{abstract}

\section{Introduction}

Recently quantum field theories on a noncommutative space-time have received
a lot of attention, see for example \cite{Obzor} and references there. The
noncommutative $d+1$ space-time can be realized by the coordinate operators $%
\hat{q}^{\mu },$ $\mu =0,1,...,d\,,$ satisfying%
\begin{equation}
\left[ \hat{q}^{\mu },\hat{q}^{\nu }\right] =i\theta ^{\mu \nu }\,,
\label{a.1}
\end{equation}%
where, in the general case, the noncommutativity parameters enter in the
theory via an antisymmetric matrix $\theta ^{\mu \nu \,}$. Obviously, many
of principle problems related to the noncommutativity can be examined
already in the noncommutative quantum mechanics (QM). Some of articles in
this direction consider generalization of the well-known QM problems
(harmonic oscillator \cite{Smalagic1}, Landau problem \cite{Nair}, Hydrogen
atom spectrum \cite{Chaichian1}, a particle in the Aharonov-Bohm field \cite%
{Chaichian}, and a system in a central potential \cite{Gamboa}) for the
noncommutative case, trying to extract possible observable differences with
the commutative case. In this connection path integral representations in
nonrelativistic QM were studied \cite{Mangano}-\cite{Tan} and calculated for
simple cases of the harmonic oscillator \cite{Acatrinei} and a free particle 
\cite{Tan}.

One ought to say that classical actions of field theories on a
noncommutative space-time can be written as some modified classical actions
already on the commutative space-time (introducing a star product). Then the
quantization of such modified actions (let us call them $\theta$-modified
actions in what follows) reproduces both space-time noncommutativity and
usual quantum mechanical features of the corresponding field theory.
Considering QM of one particle (or a system of $N$ particles) with
noncommutative coordinates, one can ask the question how to construct a $%
\theta$-modified classical action (with already commuting coordinates) for
the system. As in the case of field theory, such $\theta$-modified classical
actions in course of a quantization must reproduce both noncommutativity of
coordinates and usual QM features of the corresponding finite-dimensional
physical system. For nonrelativistic QM, the latter problem was solved by
Deriglazov in \cite{Deriglazov}. In the present article we discuss the
problem of constructing $\theta$-modified actions for relativistic QM. We
construct $\theta$-modified actions for relativistic spinless and spinning
particles. The key idea is to extract $\theta$-modified actions of the
relativistic particles from path integral representations of the
corresponding noncommtative field theory propagators. We consider $\theta$%
-modified Klein-Gordon and Dirac equations with external backgrounds for the
causal propagators. Then, using technics developed in \cite{FG,Gitman} for
usual commutative case, we construct for them path-integral representations.
Effective actions in such path-integral representations, we treat as $\theta$%
-modified actions of the relativistic particles. To confirm this
interpretation, we quantize canonically these actions. Thus, we obtain the
above mentioned $\theta$-modified Klein-Gordon and Dirac equations. The $%
\theta$-modified action of the relativistic spinning particle is a
generalization of the Berezin-Marinov pseudoclassical action \cite{Berezin}
for the noncommutative case. One ought to say that effects of the
noncommutativity appear to be essential only due to the external background.
Finally, we consider a noncommutative $d$-dimensional nonrelativistic QM
with no restrictions on the noncommutativity parameters $\theta^{\mu\nu\,}$
and formally arbitrary Hamiltonian. We construct a path integral
representation for the corresponding propagation function and demonstrate
that the effective action in our path-integral representation is just $%
\theta $-modified action for nonrelativistic QM proposed in \cite{Deriglazov}%
.

\section{Path integral representations for particle propagators in
noncommutative field theory}

\subsection{Spinless case}

In field theories the effect of the noncommutativity of the space-time can
be realized by substitution of usual function product by the Weil-Moyal star
product 
\begin{align}
f\left( x\right) \ast g\left( x\right) & =f\left( x\right) \exp\left\{ \frac{%
i}{2}\overleftarrow{\partial}_{\mu}\theta^{\mu\nu }\overrightarrow{%
\partial_{\nu}}\right\} g\left( x\right)  \notag \\
& =f\left( x^{\mu}+\frac{i}{2}\theta^{\mu\nu}\partial_{\nu}\right) g\left(
x\right) \,,  \label{13}
\end{align}
where $\ f\left( x\right) $ and $g\left( x\right) $ are two arbitrary
infinitely differentiable functions of the commutative variables $x^{\mu}$
and the second line in (\ref{13}) holds if perturbation in $\theta$ are
possible (see \cite{Chaichian}). The latter is presumably since the effect
of noncommutativity should be small.

The action of a noncommutative field theory of a scalar field $\Phi$ that
interacts with an external electromagnetic field $A_{\mu}\left( x\right) $
reads%
\begin{equation}
S_{\mathrm{scal-field}}^{\theta}=\int d^{D}x\left[ \left( P_{\mu}\ast
\Phi\right) \ast\left( P^{\mu}\ast\bar{\Phi}\right) +m^{2}\Phi\bar{\Phi }%
\right] \,,\;P_{\mu}=i\partial_{\mu}-gA_{\mu}\left( x\right) \,.  \label{15}
\end{equation}
The corresponding Euler-Lagrange equation%
\begin{equation}
\frac{\delta S_{\mathrm{scal-field}}^{\theta}}{\delta\bar{\Phi}}=0\Rightarrow%
\left[ P_{\mu}\ast P^{\mu}-m^{2}\right] \ast\Phi=0\,,  \label{16}
\end{equation}
being rewritten by the help of (\ref{13}) takes the form%
\begin{align}
& \left( \tilde{P}^{2}-m^{2}\right) \Phi=0~,\ \ \tilde{P}^{2}=\tilde {P}%
_{\mu}\tilde{P}^{\mu},  \label{16a} \\
& \tilde{P}_{\mu}=i\partial_{\mu}-gA_{\mu}\left( x^{\mu}+\frac{i}{2}%
\theta^{\mu\nu}\partial_{\nu}\right) \,,  \label{16b}
\end{align}
and is an analog of the Klein-Gordon equation for noncommutative case. The
propagator in the noncommutative scalar field theory is the causal Green
function\ $D^{c}\left( x,y\right) $ of the equation (\ref{16a}), 
\begin{equation}
\left( \tilde{P}^{2}-m^{2}\right) D^{c}\left( x,y\right) =-\delta\left(
x-y\right) ~.  \label{17}
\end{equation}

From this point on, we are going to follow the way elaborated in \cite{FG}
to construct a path integral representation for the propagator: We consider $%
D^{c}\left( x,y\right) $ as a matrix element of an operator $\hat{D}^{c}$ in
a Hilbert space%
\begin{equation}
D^{c}\left( x,y\right) =\left\langle x\right| \hat{D}^{c}\left|
y\right\rangle \,.  \label{18}
\end{equation}
Here $\left| x\right\rangle $ are eigenvectors of some self-adjoint and
mutually commuting operators $\hat{x}^{\mu},$ 
\begin{equation}
\hat{x}^{\mu}=\hat{q}^{\mu}+\frac{1}{2\hbar}\theta^{\mu\nu}\hat{p}_{\nu},
\label{cc}
\end{equation}
where operators $\hat{q}^{\mu}$ obey the commutation relations (\ref{a.1})
and $\hat{p}_{\mu}$ are momentum operators conjugated to $\hat{x}^{\mu},$%
\begin{align}
& \left[ \hat{x}^{\mu},\hat{p}_{\nu}\right] =i\hbar\delta_{\nu}^{\mu }\,,\;%
\left[ \hat{x}^{\mu},\hat{x}^{\nu}\right] =\left[ \hat{p}_{\mu},\hat{p}_{\nu}%
\right] =0\,,  \notag \\
& \hat{x}^{\mu}\left| x\right\rangle =x^{\mu}\left| x\right\rangle
\,,\;<x|y>=\delta^{D}\left( x-y\right) \,,\;\int|x><x|dx=I\,.  \label{19a}
\end{align}
Then equation (\ref{17}) implies $\hat{D}^{c}=\left( m^{2}-\Pi^{2}\right)
^{-1}$, where\footnote{%
Here and in what follows $\Pi^{2}=\Pi_{\mu}\Pi^{\mu}$ and so on.}%
\begin{align}
& \hat{\Pi}_{\mu}=-\hat{p}_{\mu}-gA_{\mu}\left( \hat{q}\right) ,\ \ \left[ 
\hat{\Pi}_{\mu},\hat{\Pi}_{\nu}\right] =-ig\hat{F}_{\mu\nu}\,,  \notag \\
& \hat{F}_{\mu\nu}=\partial_{\mu}A_{\nu}\left( \hat{q}\right)
-\partial_{\nu}A_{\mu}\left( \hat{q}\right) +ig\left[ A_{\mu}\left( \hat{q}%
\right) ,A_{\nu}\left( \hat{q}\right) \right] \,.  \label{20}
\end{align}
Due to the star product property $f\left( \hat{q}\right) g\left( \hat {q}%
\right) =\left( f\ast g\right) \left( \hat{q}\right) $, we can represent the
operator $\hat{F}_{\mu\nu}$ as follows%
\begin{equation}
\hat{F}_{\mu\nu}=\ F_{\mu\nu}^{\ast}\left( \hat{q}\right) ,\;\;F_{\mu\nu
}^{\ast}\left( q\right) =\partial_{\mu}A_{\nu}-\partial_{\nu}A_{\mu
}+ig\left( A_{\mu}\ast A_{\nu}-A_{\nu}\ast A_{\mu}\right) \,.  \label{20a}
\end{equation}

Using the Schwinger proper-time representation for the inverse operator, we
get: 
\begin{align}
& D^{c}=D^{c}\left( x_{out},x_{in}\right) =i\overset{\infty}{\underset{0}{%
\int}}\left\langle x_{out}\right\vert \exp\left[ -\frac{i}{\hbar}\hat{%
\mathcal{H}}\left( \lambda\right) \right] \left\vert x_{in}\right\rangle
d\lambda~,  \label{22} \\
& \hat{\mathcal{H}}\left( \lambda\right) =\lambda\left( m^{2}-\Pi
^{2}\right) ~.  \notag
\end{align}
Here and in what follows the infinitesimal factor $-i\epsilon$ is included
in $m^{2}$. Doing finally a discretization, similar to that in \cite{FG}, we
get a path integral representation for the propagator (\ref{22})%
\begin{equation}
D^{c}=i\overset{\infty}{\underset{0}{\int}}d\lambda_{0}\overset{x_{out}}{%
\underset{x_{in}}{\int}}Dx\underset{\lambda_{0}}{\int}D\lambda\int
DpD\pi\exp\left\{ \frac{i}{\hbar}\left[ S_{\mathrm{scal-part}}^{\theta }+S_{%
\mathrm{GF}}\right] \right\} ,  \label{23}
\end{equation}
where%
\begin{align}
& S_{\mathrm{scal-part}}^{\theta}=\overset{1}{\underset{0}{\int}}\left[
\lambda\left( \mathcal{P}^{2}-m^{2}\right) +p_{\mu}\dot{x}^{\mu}\right]
d\tau\,,\;S_{\mathrm{GF}}=\overset{1}{\underset{0}{\int}}\pi\dot{\lambda}%
d\tau\,,  \notag \\
& \mathcal{P}_{\mu}=-p_{\mu}-gA_{\mu}\left( x^{\mu}-\frac{1}{2\hbar}%
\theta^{\mu\nu}p_{\nu}\right) ,\;\dot{x}=\frac{dx}{d\tau}\,,\;\dot{\lambda }=%
\frac{d\lambda}{d\tau}\,.  \label{23a}
\end{align}
The functional integration in (\ref{23}) goes over trajectories $x^{\mu
}\left( \tau\right) ,~p_{\mu}\left( \tau\right) ,~\lambda\left( \tau\right)
, $ and $\pi\left( \tau\right) ,$ parametrized by some invariant parameter $%
\tau\in\left[ 0,1\right] $ and obeying the boundary conditions $x\left(
0\right) =x_{in},~x\left( 1\right) =x_{out},~\lambda\left( 0\right)
=\lambda_{0}$.

Since momenta are involved in arguments of electromagnetic potentials $%
A_{\mu }$, an integration over the momenta in the representation (\ref{23})
is difficult to perform in the general case. On the other side, we can go
over from $x$ to new\ coordinates $q,$%
\begin{equation}
q^{\mu}=x^{\mu}-\frac{1}{2\hbar}\theta^{\mu\nu}p_{\nu}~,  \label{24}
\end{equation}
which correspond in a sense to the noncommutative operators $\hat{q}^{\mu}$ (%
\ref{a.1}). Then 
\begin{align}
& D^{c}=i\overset{\infty}{\underset{0}{\int}}d\lambda_{0}\overset{%
x_{out}-\theta p/2\hbar}{\underset{x_{in}-\theta p/2\hbar}{\int}}Dq\underset{%
\lambda_{0}}{\int}D\lambda\int DpD\pi\exp\left\{ \frac{i}{\hbar }S_{\mathrm{%
scal-part}}^{\theta}+S_{\mathrm{GF}}\right\} \,,  \notag \\
& S_{\mathrm{scal-part}}^{\theta}=\overset{1}{\underset{0}{\int}}\left\{
\lambda\left[ \left( p_{\mu}+gA_{\mu}\left( q\right) \right) ^{2}-m^{2}%
\right] +p_{\mu}\dot{q}^{\mu}+\frac{1}{2\hbar}\dot{p}_{\mu}\theta
^{\mu\nu}p_{\nu}\right\} d\tau.  \label{25}
\end{align}
Thus, we get rid from the above mentioned difficulty but a new one has
appeared. The action $S_{\mathrm{scal}}^{\theta}$ in (\ref{25}) contains an
\textquotedblright inconvenient\textquotedblright\ term $\dot{p}%
_{\mu}\theta^{\mu\nu}p_{\nu}/2\hbar$. Here a possibility to integrate over
momenta is related to the study of the structure of $\theta^{\mu\nu}$ matrix
and with a subsequent transition to some Darboux coordinates.

The representation (\ref{25}) can be treated as a Hamiltonian path integral
for the scalar particle propagator in the noncommutative field theory. The
exponent in the integrand (\ref{25}) can be considered as an effective and
non-degenerate Hamiltonian action of a scalar particle in a noncommutative
space time. It consists of two parts. The first one $S_{\mathrm{GF}}$ can be
treated as a gauge fixing term and corresponds, in fact, to the gauge
condition $\dot{\lambda}=0$. The rest part of the effective action $S_{%
\mathrm{scal-part}}^{\theta}$ can be treated as $\theta$-modification of the
usual Hamiltonian action of a spinless relativistic particle in the
commutative case. This action differs from the corresponding commutative
case (see \cite{FG}) by the term $\frac{1}{2\hbar}\dot{p}_{\mu}\theta^{\mu%
\nu }p_{\nu}$ .

\subsection{Spinning particle}

Consider a $\theta$-modified action of noncommutative field theory of a
spinor field $\Psi$ that interacts with an external electromagnetic
background $A_{\mu}.$ Being written in commuting $D$-dimensional Minkowski
coordinates $x^{\mu},\;\mu=0,1,...,D-1,$ the action reads%
\begin{equation}
S_{\mathrm{spinor-field}}^{\theta}=\int dx^{D}\bar{\Psi}\ast\left( P_{\mu
}\gamma^{\mu}+m\right) \ast\Psi~,  \label{27}
\end{equation}
where $\gamma^{\mu}$ are gamma-matrices in $D$ dimensions, $\left[
\gamma^{\mu},\gamma^{\nu}\right] _{+}=2\eta^{\mu\nu}\;.$ In this article, we
consider $D$ to be even $D=2d,$ for simplicity and as a generalization of
the $4$-dim. Minkowski space, the odd case can be considered in the same
manner following ideas of the work \cite{Gitman}. As it is known \cite{BW},
in even dimensions a matrix representation of the Clifford algebra with
dimensionality $\mathrm{dim}\,\gamma^{\mu}=2^{d}$ always exists. In other
words ${\gamma ^{\mu}}$ are $2^{d}\times2^{d}$ matrices. In such dimensions
one can introduce another matrix, $\gamma^{D+1}=r\gamma^{0}\gamma^{1}\ldots%
\gamma^{D-1},$ where $r=1$,\ if$\;d$\ is\ even, and $r=i,\;$if\ d\ is\ odd,
which anticommutes with all $\gamma^{\mu}$ (analog of $\gamma^{5}$ in four
dimensions), $[\gamma ^{D+1},\gamma^{\mu}]_{+}=0,\;\left(
\gamma^{D+1}\right) ^{2}=-1.$ The Euler-Lagrange equations%
\begin{equation}
\frac{\delta S_{\mathrm{spinor-field}}^{\theta}}{\delta\bar{\Psi}}%
=0\Rightarrow\left( P_{\mu}\gamma^{\mu}+m\right) \ast\Psi=0~,  \label{28}
\end{equation}
beeing rewritten by the help of (\ref{13}) take the form%
\begin{equation}
\left( \tilde{P}_{\mu}\gamma^{\mu}-m\right) \Psi=0\,,\;\tilde{P}_{\mu
}=i\partial_{\mu}-gA_{\mu}\left( x^{\mu}+\frac{i}{2}\theta^{\mu\nu}%
\partial_{\nu}\right) \,,  \label{28a}
\end{equation}
and represent an analog of the Dirac equation for the noncommutative case.
The propagator of the noncommutative spinor field theory is the causal Green
function\ $G^{c}\left( x,y\right) $ of equation (\ref{28a}),%
\begin{equation}
\left( \tilde{P}_{\mu}\gamma^{\mu}-m\right) G^{c}(x,y)=-\delta^{D}(x-y)\,.
\label{b1}
\end{equation}

Following \cite{FG,Gitman}, we pass to a $\theta$-modified Dirac operator
which is homogeneous in $\gamma$-matrices. Indeed, let us rewrite the
equation (\ref{b1}) in terms of the transformed by $\gamma^{D+1}$ propagator 
$\tilde {G}^{c}(x,y)$, 
\begin{equation}
\tilde{G}^{c}(x,y)=G^{c}(x,y)\gamma^{D+1},\;\;\left( \tilde{P}_{\mu}\tilde{%
\gamma}^{\mu}-m\gamma^{D+1}\right) \tilde{G}^{c}(x,y)=\delta^{D}(x-y),
\label{b3}
\end{equation}
where $\tilde{\gamma}^{\mu}=\gamma^{D+1}\gamma^{\mu}$. The matrices $\tilde{%
\gamma}^{\mu}$ have the same commutation relations as initial ones without
tilda $\left[ \tilde{\gamma}^{\mu},\tilde{\gamma}^{\nu}\right]
_{+}=2\eta^{\mu\nu}$, and anticommute with the matrix $\gamma^{D+1}$. The
set of $D+1$ gamma-matrices $\tilde{\gamma}^{\nu}$ and $\gamma^{D+1}$ form a
representation of the Clifford algebra in odd $2d+1$ dimensions. Let us
denote such matrices via $\Gamma^{n}$, 
\begin{align}
& \Gamma^{n}=\left\{ 
\begin{array}{ll}
\tilde{\gamma}^{\mu}, & n=\mu=0,\ldots,D-1 \\ 
\gamma^{D+1}, & n=D%
\end{array}
\right. \;,  \label{b4} \\
& [\Gamma^{k},\Gamma^{n}]_{+}=2\eta^{kn},\;\;\eta_{kn}=\mathrm{diag}(%
\underbrace{1,-1,\ldots,-1}_{D+1}),\;k,n=0,\ldots,D\,.  \notag
\end{align}
In terms of these matrices the equation (\ref{b3}) takes the form 
\begin{equation}
\tilde{P}_{n}\Gamma^{n}\tilde{G}^{c}(x,y)=\delta^{D}(x-y),\;\;\tilde{P}_{\mu
}=i\partial_{\mu}-gA_{\mu}\left( x^{\mu}+\frac{i}{2}\theta^{\mu\nu}%
\partial_{\nu}\right) ,\;\;\tilde{P}_{D}=-m\;.  \label{b5}
\end{equation}
Now again, similar to (\ref{18}), we present $\tilde{G}^{c}(x,y)$ as a
matrix element of an operator $\hat{G}^{c}$ (in the\ coordinate
representation (\ref{19a})), 
\begin{equation}
\tilde{G}_{ab}^{c}(x,y)=<x|\hat{G}_{ab}^{c}|y>,\;\;a,b=1,2,\ldots,2^{d}\,,
\label{b6}
\end{equation}
where the spinor indices $a,b$ are written here explicitly for clarity and
will be omitted hereafter. The equation (\ref{b5}) implies $\hat{S}%
^{c}=\left( \Pi_{n}\Gamma^{n}\right) ^{-1},$ where $\Pi_{\mu}$ are defined
in (\ref{20}), and $\Pi_{D}=-m$. Using a generalization of the Schwinger
proper-time representation, proposed in \cite{FG}, we write the Green
function (\ref{b6}) in the form 
\begin{align}
& \tilde{G}^{c}=\tilde{G}^{c}(x_{out},x_{in})=\int_{0}^{\infty}\,d\lambda
\int\langle x_{\mathrm{out}}|e^{-i\hat{\mathcal{H}}(\lambda,\chi )}|x_{%
\mathrm{in}}\rangle d\chi\,,  \label{b10} \\
& \hat{\mathcal{H}}(\lambda,\chi)=\lambda\left( m^{2}-\Pi^{2}+\frac{ig}{2}%
F_{\mu\nu}\Gamma^{\mu}\Gamma^{\nu}\right) +\Pi_{n}\Gamma^{n}\,\chi\;.  \notag
\end{align}

Similar to \cite{FG}, we present the matrix element entering in the
expression (\ref{b10}) by means of a Hamiltonian path integral%
\begin{align}
& \tilde{G}^{c}=\exp\left( i\Gamma^{n}\frac{\partial_{l}}{\partial
\varepsilon^{n}}\right) \int_{0}^{\infty}\,d\lambda_{0}\int
d\chi_{0}\int_{\lambda_{0}}D\lambda\int_{\chi_{0}}D\chi%
\int_{x_{in}}^{x_{out}}Dx\int Dp\int D\pi\int D\nu  \label{b15a} \\
& \times\int_{\psi(0)+\psi(1)=\varepsilon}\mathcal{D}\psi\exp\left\{
i\int_{0}^{1}\left[ \lambda\left( \mathcal{P}^{2}-m^{2}+2igF_{\mu\nu}^{\ast
}\psi^{\mu}\psi^{\nu}\right) +2i\mathcal{P}_{n}\psi^{n}\chi\right. \right. 
\notag \\
& -i\psi_{n}\dot{\psi}^{n}+\left. \left. p_{\mu}\dot{x}^{\mu}+\pi \dot{%
\lambda}+\nu\dot{\chi}\right] d\tau+\left. \psi_{n}(1)\psi ^{n}(0)\right\}
\right\vert _{\varepsilon=0}\,.  \notag
\end{align}
Here $\varepsilon^{n}$ are odd variables, anticommuting with the $\Gamma $%
-matrices,%
\begin{equation}
\mathcal{P}_{\mu}=-p_{\mu}-gA_{\mu}\left( x^{\mu}-\frac{1}{2\hbar}\theta
^{\mu\nu}p_{\nu}\right) ,\;\mathcal{P}_{D}=-m,\;F_{\mu\nu}^{\ast}=F_{\mu\nu
}^{\ast}\left( x^{\mu}-\frac{1}{2\hbar}\theta^{\mu\nu}p_{\nu}\right) , 
\notag
\end{equation}
the function $F_{\mu\nu}^{\ast}\left( q\right) $ is defined in (\ref{20a}),\
and the integration goes over even trajectories $x\left( \tau\right)
,~p\left( \tau\right) ,~\lambda\left( \tau\right) ,$ $\pi\left( \tau\right)
, $ and odd trajectories $\psi_{n}(\tau),$ $\chi (\tau),\;\nu(\tau),$
parametrized by some invariant parameter $\tau\in\left[ 0,1\right] $ and
obeying the boundary conditions $\,x(0)=x_{\mathrm{in}}$, $x(1)=x_{\mathrm{%
out}}$, $\lambda(0)=\lambda_{0}$, $\chi(0)=\chi_{0}$.

Performing the change of variables (\ref{24})$\ $in (\ref{b15a}), we obtain
another representaion for $\tilde{G}^{c},$%
\begin{align}
& \tilde{G}^{c}=\exp \left( i\Gamma ^{n}\frac{\partial _{l}}{\partial
\varepsilon ^{n}}\right) \int_{0}^{\infty }\,d\lambda _{0}\int d\chi
_{0}\int_{\lambda _{0}}D\lambda \int_{\chi _{0}}D\chi \overset{\infty }{%
\underset{-\infty }{\int }}Dp\overset{x_{out}-\theta p/2\hbar }{\underset{%
x_{in}-\theta p/2\hbar }{\int }}Dq\int D\pi \int D\nu   \label{b16} \\
& \times \int_{\psi (0)+\psi (1)=\varepsilon }\mathcal{D}\psi \left. \exp
\left\{ i\left[ S_{\mathrm{spin-part}}^{\theta }+S_{\mathrm{GF}}\right]
+\psi _{n}(1)\psi ^{n}(0)\right\} \right\vert _{\varepsilon =0}\,,  \notag
\end{align}%
where 
\begin{subequations}
\begin{align}
& S_{\mathrm{spin-part}}^{\theta }=\int_{0}^{1}\left[ \lambda \left( \left(
p_{\mu }+gA_{\mu }\right) ^{2}-m^{2}+2igF_{\mu \nu }^{\ast }\psi ^{\mu }\psi
^{\nu }\right) +2i\left( p_{\mu }+gA_{\mu }\left( q\right) \right) \psi
^{\mu }\chi \right.   \notag \\
& \left. -2im\psi ^{D}\chi -i\psi _{n}\dot{\psi}^{n}+p_{\mu }\dot{q}^{\mu }+%
\frac{1}{2\hbar }\dot{p}_{\mu }\theta ^{\mu \nu }p_{\nu }\right] d\tau \;,
\label{BM} \\
& S_{\mathrm{GF}}=\int_{0}^{1}\left( \pi \dot{\lambda}+\nu \dot{\chi}\right)
d\,\tau .  \label{GF}
\end{align}

Note that in the work \cite{BB} was made an attempt to construct the path
integral representation of Green function of noncommutative Dirac equation.
However the consideration was perturbative in $\theta $ (taken into account
only the first order perturbation). As a consequence the authors did not
obtain the corresponding action (\ref{BM}), moreover the essential term  $%
\dot{p}_{\mu }\theta ^{\mu \nu }p_{\nu }/2\hbar $ was missed.

\section{Pseudoclassical action of spinning particle in noncommutative space
time}

Similar to the spinless case, the exponent in the integrand (\ref{b16}) can
be considered as an effective and non-degenerate Hamiltonian action of a
spinning particle in the noncommutative space time. It consists of two
principal parts. The first one $S_{\mathrm{GF}}$ with derivatives of $%
\lambda $ and $\chi$ can be treated as a gauge fixing term, which
corresponds to gauge conditions $\dot{\lambda}=\dot{\chi}=0$. The rest part $%
S_{\mathrm{spin-part}}^{\theta}$ can be treated as a gauge invariant action
of a spinning particle in the noncommutative space time. The action $S_{%
\mathrm{spin-part}}^{\theta}$ is a $\theta$-modification of the Hamiltonian
form of the Berezin-Marinov action \cite{Berezin}. It will be studied and
quantized below to justify such an interpretation.

One can easily verify that $S_{\mathrm{spin-part}}^{\theta}$ is
reparametrization invariant. Explicit form of supersymmetry transformations,
which generalize ones for the Berezin-Marinov action, is not so easily to
derive. Their presence will be proved in an indirect way. Namely, we are
going to prove the existence of two primary first-class constraints in the
corresponding Hamiltonian formulation.

Let us consider $S_{\mathrm{spin-part}}^{\theta}$ as a Lagrangian action
with generalized coordinates $Q_{A}=\left( q^{\mu},p_{\mu}\right) $, $%
A=(\zeta,\mu),$ $\zeta=1,2,\;Q_{1\mu}=q^{\mu},\;Q_{2\mu}=p_{\mu}$ ; $\chi,$ $%
\psi,$ and $\lambda,$ and let us perform a Hamiltonization of such an
action. To this end, we introduce the canonical momenta $P$ conjugate to the
generalized coordinates as follows: 
\end{subequations}
\begin{align}
& P_{Q_{A}}=\frac{\partial L}{\partial\dot{Q}^{A}}=J_{A}\left( Q\right)
\,,\;J_{1\mu}=p_{\mu}\,,\ \ J_{2\mu}=\frac{1}{2\hbar}\theta^{\mu\nu}p_{\nu
}\,,  \notag \\
& P_{\lambda}=\frac{\partial L}{\partial\dot{\lambda}}=0,\;P_{\chi }=\frac{%
\partial_{r}L}{\partial\dot{\chi}}=0,\;P_{n}=\frac{\partial_{r}L}{\partial%
\dot{\psi}^{n}}=-i\psi_{n}\,.  \label{d1}
\end{align}
\break It follows from equations (\ref{d1}) that there exist primary
constraints $\Phi^{(1)}=0$, 
\begin{equation}
\Phi_{l}^{(1)}=\left\{ 
\begin{array}{l}
\Phi_{1A}^{(1)}=P_{A}-J_{A}\left( Q\right) ~, \\ 
\Phi_{2}^{(1)}=P_{\lambda}\,\,\,,\ \ \Phi_{3}^{\left( 1\right) }=P_{\chi }~,
\\ 
\Phi_{4n}^{(1)}=P_{n}+i\psi_{n}\,\,\,.%
\end{array}
\right.  \label{d2}
\end{equation}
The Poisson brackets of primary constraints are%
\begin{align}
& \{\Phi_{1A}^{\left( 1\right) },\Phi_{1B}^{\left( 1\right)
}\}=\Omega_{AB}=\left( 
\begin{array}{cc}
\mathbf{0} & \mathbb{I} \\ 
-\mathbb{I} & \mathbf{\theta}/\hbar%
\end{array}
\right) \,,\;\left\{ \Phi_{4n}^{(1)},\Phi_{4m}^{(1)}\right\} =2i\eta _{nm}~,
\\
& \{\Phi_{1A}^{\left( 1\right) },\Phi_{4n}^{\left( 1\right)
}\}=\{\Phi_{1A}^{\left( 1\right) },\Phi_{2,3}^{\left( 1\right)
}\}=\{\Phi_{4n}^{\left( 1\right) },\Phi_{2,3}^{\left( 1\right) }\}=0~. 
\notag
\end{align}
where $\mathbf{\theta}=\theta^{\mu\nu\,},$\ $\mathbb{I}$ is a $D\times D$
unit matrix, and $\mathbf{0}$ denotes an $D\times D$ zero matrix. Note that $%
\det\Omega_{AB}=1$, and 
\begin{equation*}
\omega^{AB}=\Omega_{AB}^{-1}=\left( 
\begin{array}{cc}
\mathbf{\theta}/\hbar & -\mathbb{I} \\ 
\mathbb{I} & \mathbf{0}%
\end{array}
\right) ~.
\end{equation*}

Now we construct the total Hamiltonian $H^{(1)}$, according to the standard
procedure \cite{GTbook}. Thus, we obtain: 
\begin{align}
& H^{(1)}=H+\Lambda_{l}\Phi_{l}^{(1)},  \notag \\
& H=-\lambda\left[ \left( p_{\mu}+gA_{\mu}\right) ^{2}-m^{2}+2igF_{\mu\nu
}^{\ast}\left( q\right) \psi^{\mu}\psi^{\nu}\right] +2i\chi\left( \left(
p_{\mu}+gA_{\mu}\right) \psi^{\mu}-m\psi^{D}\,\right) \,.  \label{d3}
\end{align}
where $\Lambda_{l}$\ .... The consistency conditions $\dot{\Phi}%
_{1A,4n}^{(1)}=\left\{ {\Phi}_{1A,4n}^{(1)},H^{(1)}\right\} =0$ for the
primary constraints $\Phi_{1A}^{(1)}$ and $\Phi_{4n}^{(1)}$ allow us to fix
the Lagrange multipliers $\lambda^{1A}$ and $\lambda^{4n}$. The consistency
conditions for the constraints $\Phi_{2,3}^{(1)}$ imply secondary
constraints $\Phi_{1,2}^{(2)}=0$, 
\begin{align}
& \Phi_{1}^{(2)}=\left( p_{\mu}+gA_{\mu}\right) \psi^{\mu}-m\psi ^{D}=0\,,
\label{d4} \\
& \Phi_{2}^{(2)}=\left( p_{\mu}+gA_{\mu}\right) ^{2}-m^{2}+2igF_{\mu\nu
}^{\ast}\psi^{\mu}\psi^{\nu}=0\,.  \label{d5}
\end{align}
Thus, the Hamiltonian $H$ appears to be proportional to constraints, as
always in the case of a repara\-metrization invariant theory, 
\begin{equation*}
H=2i\chi\Phi_{1}^{(2)}-\lambda\Phi_{2}^{(2)}.
\end{equation*}
No more secondary constraints arise from the Dirac procedure, and the
Lagrange multipliers $\lambda^{2}$ and $\lambda^{3}$ remain undetermined, in
perfect correspondence with the fact that the number of gauge
transformations parameters equals two for the theory in question.

One can go over from the initial set of constraints $\left(
\Phi^{(1)},\Phi^{(2)}\right) $ to the equivalent one $\left(
\Phi^{(1)},T\right) ,$ where: 
\begin{equation}
T=\Phi^{(2)}+\frac{\partial\Phi^{\left( 2\right) }}{\partial q^{A}}%
\omega^{AB}\Phi_{1B}^{(1)}+\frac{i}{2}\frac{\partial_{r}\Phi}{\partial\psi
^{n}}^{(2)}\Phi_{4n}^{(1)}\,.  \label{d6}
\end{equation}
The new set of constraints can be explicitly divided in a set of first-class
constraints, which is $\left( \Phi_{2,3}^{(1)},T\right) $ and in a set of
second-class constraints, which is $\left(
\Phi_{1A}^{(1)},\Phi_{4n}^{(1)}\right) $.

Now we consider an operator quantization. To this end we perform a partial
gauge fixing, imposing gauge conditions $\Phi_{1,2}^{\mathrm{G}}=0$ to the
primary first-class constraints $\Phi_{1,2}^{(1)}\,$, 
\begin{equation}
\Phi_{1}^{\mathrm{G}}=\chi=0,\,\,\,\,\,\,\Phi_{2}^{\mathrm{G}}=\lambda=1/m\,.
\label{d7}
\end{equation}
One can check that the consistency conditions for the gauge conditions (\ref%
{d7}) lead to fixing the Lagrange multipliers $\lambda_{2}$ and $\lambda_{3}$%
. Thus, on this stage we reduced our Hamiltonian theory to one with the
first-class constraints $T$ and second-class ones $\varphi=\left(
\Phi^{(1)},\Phi^{\mathrm{G}}\right) $. Then, we apply the so called Dirac
method for systems with first-class constraints \cite{Dirac64}, which, being
generalized to the presence of second-class constraints, can be formulated
as follow: the commutation relations between operators are calculated
according to the Dirac brackets with respect to the second-class constraints
only; second-class constraints as operators equal zero; first-class
constraints as operators are not zero, but, are considered in sense of
restrictions on state vectors. All the operator equations have to be
realized in a Hilbert space.

The subset of the second-class constraints $\left( \Phi_{2,3}^{(1)},\Phi^{%
\mathrm{G}}\right) $ has a special form \cite{GTbook}, so that one can use
it for eliminating of the variables $\lambda,P_{\lambda},\chi,P_{\chi}$,
from the consideration, then, for the rest of the variables $q,p,\psi^{n}$,
the Dirac brackets with respect to the constraints $\varphi$ reduce to ones
with respect to the constraints $\Phi_{1A}^{(1)}$ and $\Phi_{4n}^{(1)}$ only
and can be easy calculated, 
\begin{equation*}
\left\{ Q^{A},Q^{B}\right\}
_{D(\Phi^{(1)})}=\omega^{AB}\,,\,\,\,\,\,\,\,\left\{
\psi^{n},\psi^{m}\right\} _{D(\Phi^{(1)})}=\frac{i}{2}\eta^{nm}\,,
\end{equation*}
while all other Dirac brackets vanish. Thus, the commutation relations for
the operators $\hat{q}^{\mu},\hat{p}_{\mu},\hat{\psi}^{n}$, which correspond
to the variables $q^{\mu},p_{\mu},\psi^{n}$ respectively, are 
\begin{align}
& \left[ \hat{q}^{\mu},\hat{p}_{\nu}\right] _{-}=i\hbar\omega^{\mu,D+\nu
}=i\hbar\delta_{\nu}^{\mu}\,,\;\left[ \hat{q}^{\mu},\hat{q}^{\nu}\right]
=i\hbar\omega^{\mu\nu}=i\theta^{\mu\nu},\;\left[ \hat{p}_{\mu},\hat{p}_{\nu }%
\right] =0,  \notag \\
& \left[ \hat{\psi}^{m},\hat{\psi}^{n}\right] _{+}=i\left\{
\psi^{m},\psi^{n}\right\} _{D(\Phi^{(1)})}=-\frac{1}{2}\eta^{mn}\,.
\label{d8}
\end{align}
Besides, the following operator equations hold: 
\begin{equation}
\hat{\Phi}_{1A}^{(1)}=\hat{P}_{A}-J_{A}\left( \hat{Q}\right) ,\;\hat{\Phi }%
_{4n}^{(1)}=\hat{P}_{n}+i\hat{\psi}_{n}=0.  \label{d9}
\end{equation}
Taking that into account, one can construct a realization of the commutation
relations (\ref{d8}) in a Hilbert space whose elements $\Psi$ are $2^{d}$%
-component columns dependent only on $x$, such that 
\begin{equation}
\hat{q}^{\mu}=\left( x^{\mu}+\frac{i}{2}\theta^{\mu\nu}\partial_{\nu}\right) 
\mathbf{I}\,,\;\;\hat{p}_{\mu}=-i\partial_{\mu}\mathbf{I}\,,\;\;\hat{\psi}%
^{n}=\frac{i}{2}\Gamma^{n}\,,  \label{c13}
\end{equation}
where $\mathbf{I}$ is $2^{d}\times2^{d}$ unit matrix, and $\Gamma^{n}$, are
gamma-matrices (\ref{b4}). The first-class constraints $\hat{T}$ as
operators have to annihilate physical vectors; in virtue of (\ref{d9}) and (%
\ref{d6}) that implies the equations: 
\begin{equation}
\hat{\Phi}_{1}^{(2)}\Psi=0\,,\;\;\hat{\Phi}_{2}^{(2)}\Psi=0\,,  \label{d10}
\end{equation}
where $\hat{\Phi}_{1,2}^{(2)}$ are operators, which correspond to
constraints (\ref{d4}), (\ref{d5}). Taking into account the realizations of
the commutation relations (\ref{d8}), one easily can see that the first
equation (\ref{d10}) takes the form of the $\theta$-modified Dirac equation, 
\begin{equation}
\left( \tilde{P}_{\mu}\tilde{\gamma}^{\mu}-m\gamma^{D+1}\right)
\Psi=0\Longleftrightarrow\left( P_{\mu}\gamma^{\mu}+m\right) \ast\Psi=0~,
\label{d11}
\end{equation}

Since $\hat{\Phi}_{2}^{(2)}=\left( \hat{\Phi}_{1}^{(2)}\right) ^{2}$, the
second equation (\ref{d10}) is a consequence of the first one.

Thus, we have constructed a $\theta$-modification of the Berezin-Marinov
action (\ref{BM}) which, being quantized, leads to a quantum theory based on
the $\theta$-modified Dirac equation.

Note that space-time non-commutativity $\left[ \hat{q}^{0},\hat{q}^{i}\right]
=i\theta^{0i}$ can be obtained also from the canonical quantization of the
conventional Lagrangian action of a relativistic spinless particle by
imposing a special gauge condition $\Phi_{gf}=x^{0}+\theta
^{0i}p_{i}-\tau=0, $ see \cite{PS}.

\section{Path integral in nonrelativistic quantum mechanics on a
noncommutative space}

In this section, we construct a path integral representation for the
propagation function (a symbol of the evolution operator) in nonrelativistic
QM on a noncommutative space. We compare our result with some previous
constructions and use it to extract a $\theta$-modified first-order
classical Hamiltonian action for such a system.

We consider a $d$-dimensional nonrelativistic QM with basic canonical
operators of coordinates $\hat{q}^{k}$ and momenta $\hat{p}_{j},$ $%
k,j=1,...,d$ that obey the following commutation relations 
\begin{equation}
\left[ \hat{q}^{k},\hat{q}^{j}\right] =i\theta^{kj}\,,\;\left[ \hat{q}^{k},%
\hat{p}_{j}\right] =i\hbar\delta_{j}^{k}\,,\;\left[ \hat{p}_{k},\hat {p}_{j}%
\right] =0\,.  \label{1}
\end{equation}
It is supposed that the nonzeroth commutation relations for the coordinate
operators in (\ref{1}) have emerged from the noncommutative properties of
the position space. The time evolution of the system under consideration is
governed by a self-adjoint Hamiltoniam $\hat{H}.$ We believe that behind
such a QM there exist a classical theory with a $\theta$-modified action
(which we are going to restore in what follows), such that a quantization of
this action leads to the QM.

In conventional nonrelativistic QM, one constructs a path integral
representations for matrix elements (in a coordinate representation) of the
evolution operator $\hat{U}\left( t,t^{\prime}\right) .$ In the QM under
consideration, we also start with such an operator. It obeys the Schr\"{o}%
dinger equation and for time independent $\hat{H}$ (which we consider for
simplicity in what follows) has the form%
\begin{equation}
\hat{U}\left( t^{\prime},t\right) =\exp\left\{ -\frac{i}{\hbar }\hat{H}%
\left( t^{\prime}-t\right) \right\} \,.  \label{2}
\end{equation}

Since the coordinate operators $\hat{q}$ do not commute, they do not posses
a common complete set of eigenvectors. Therefore, there is no $q$-coordinate
representation and one cannot speak about matrix elements of the evolution
operator in such a representation. Consequently, one cannot define a
probability amplitude of a transition between two points in the position
space. Nevertheless, one can consider another types of matrix elements of
the evolution operator that are probability amplitudes (evolution functions)
and can be represented via path integrals. Below, we consider two types of
such matrix elements, 
\begin{equation}
G_{p}=\left\langle p^{out}\right| \hat{U}\left( t_{out},t_{in}\right) \left|
p^{in}\right\rangle \;\mathrm{and\;}G_{x}=\left\langle x_{out}\right| \hat{U}%
\left( t_{out},t_{in}\right) \left| x_{in}\right\rangle \,.  \label{2a}
\end{equation}
In (\ref{2a}) $\left| p\right\rangle $ is a complete set of eigenvectors of
commuting operators $\hat{p},$%
\begin{align}
& \hat{p}_{j}\left| p\right\rangle =p_{j}\left| p\right\rangle
\,,\;<p|p^{\prime}>=\delta\left( p-p^{\prime}\right) \,,\;\int
|p><p|dp=I\,,\;dp=\prod_{i}dp_{i}\,,  \notag \\
& <p|x>=\frac{1}{\left( 2\pi\hbar\right) ^{d/2}}\exp\left\{ -\frac{i}{\hbar}%
p_{i}x^{i}\right\} \,,\;<p|\hat{x}|p^{\prime}>=i\hbar\frac{\partial }{%
\partial p}<p|p^{\prime}>\,,  \label{3}
\end{align}
and $\left| x\right\rangle $ is a complete set of eigenvectors of some
commuting and canonically conjugated to $\hat{p}$ operators $\hat{x}^{k}.$
We chose these operators as follows\footnote{%
For the first time the commuting operators $\hat{x}^{k}$ were introduced in 
\cite{Chaichian1}.}:%
\begin{align}
& \hat{x}^{k}=\hat{q}^{k}+\frac{\theta^{kj}\hat{p}_{j}}{2\hbar}\,,\;\left[ 
\hat{x}^{k},\hat{x}^{j}\right] =0\,,\;\left[ \hat{x}^{k},\hat{p}_{j}\right]
=i\hbar\delta_{j}^{k}\,,  \notag \\
& \hat{x}^{\mu}\left| x\right\rangle =x^{\mu}\left| x\right\rangle
\,,\;<x|y>=\delta^{D}\left( x-y\right) \,,\;\int|x><x|dx=I\,,\;dx=\prod
_{i}dx^{i}\,.  \label{4a}
\end{align}

First, let us construct a path integral representation for the evolution
function $G_{p}.$ To this end, as usual, we devide the time interval $%
T=t_{out}-t_{in}$ in $N$ equal parts $\Delta t=T/N$ by means of the points $%
t_{k}$, $k=1...N-1,$ such that$\ t_{k}=t_{in}+k\Delta t$. Using the group
property of the evolution operator and the completeness relation (see (\ref%
{3})) for the set $\left\vert p\right\rangle $, one can write 
\begin{equation}
G_{p}=\lim_{N\rightarrow\infty}\overset{\infty}{\underset{-\infty}{\int}}%
dp^{\left( 1\right) }...dp^{\left( N-1\right) }\prod_{k=1}^{N}<p^{\left(
k\right) }|\exp\left\{ -\frac{i}{\hbar}\hat{H}\left( t_{k}-t_{k-1}\right)
\right\} |p^{\left( k-1\right) }>\,,  \label{6}
\end{equation}
where $p^{\left( 0\right) }=p^{\left( in\right) }$, $p^{\left( N\right)
}=p^{\left( out\right) }$, and $p^{\left( k\right) }=(p_{i}^{\left( k\right)
}).$ Bearing in mind the limiting process $N\rightarrow\infty$ or $\Delta
t\rightarrow0$ and using the completeness relation (\ref{4a}) for the
eigenvectors $\left\vert x\right\rangle $, one can approximately calculate
the matrix element from (\ref{6}),%
\begin{equation}
<p^{\left( k\right) }|\exp\left\{ -\frac{i}{\hbar}\hat{H}\Delta t\right\}
|p^{\left( k-1\right) }>\thickapprox\int dx_{(k)}<p^{\left( k\right) }|1-%
\frac{i}{\hbar}\hat{H}\Delta t|x_{(k)}><x_{(k)}|p^{\left( k-1\right) }>\,,
\label{7}
\end{equation}
where $x_{(k)}=\left( x_{(k)}^{i}\right) $ and $dx_{(k)}=%
\prod_{i}dx_{(k)}^{i}.$ A result of this calculation can be expressed in
terms of a classical Hamiltonian $H$, however, in general case, it will
depend on the choice of the correspondance rule between the classical
function and quantum operator. For our calculations we choose the Weyl
ordering. In this case the matrix element (\ref{7}) will take the form%
\begin{equation*}
\int\frac{dx_{(k)}}{\left( 2\pi\hbar\right) ^{d}}\exp\left\{ \frac{i}{\hbar}%
\left[ -x_{(k)}^{i}\frac{p_{i}^{\left( k\right) }-p_{i}^{\left( k-1\right) }%
}{\Delta t}-H\left( x_{\left( k\right) }-\frac{\mathbf{\theta }p^{\left(
k\right) \prime}}{2\hbar},p^{\left( k\right) \prime}\right) \right] \Delta
t+O\left( \Delta t^{2}\right) \right\} ,
\end{equation*}
where $p^{\left( k\right) \prime}=\frac{p^{\left( k\right) }+p^{\left(
k-1\right) }}{2},$ and $H\left( x-\frac{\mathbf{\theta}p}{2\hbar},p\right) $
is the Weyl symbol of the operator $\hat{H}.$ Using the above formula and
taking the limit $N\rightarrow\infty$ in the integral (\ref{6}), we get for $%
G_{p}$ the following path integral representation:%
\begin{equation}
G_{p}=\overset{p^{\left( out\right) }}{\underset{p^{\left( in\right) }}{\int}%
}Dp\int Dx\exp\left\{ \frac{i}{\hbar}\int dt\left[ -x_{j}\dot{p}^{j}-H\left(
x-\frac{\mathbf{\theta}p}{2\hbar},p\right) \right] \right\} \,.  \label{8}
\end{equation}

In the same manner, one can construct a path integral representation for the
evolution function $G_{x},$ which, is%
\begin{equation}
G_{x}=\int Dp\int_{x_{(in)}}^{x_{(out)}}Dx\exp\left\{ \frac{i}{\hbar}\int dt%
\left[ p_{j}\dot{x}^{j}-H\left( x-\frac{\mathbf{\theta}p}{2\hbar },p\right) %
\right] \right\} \,.  \label{9}
\end{equation}

Let us pass to the integration over trajectories $q=x-\frac{\mathbf{\theta}p%
}{2\hbar}$ in path integrals (\ref{8}) and (\ref{9}). Then we get%
\begin{align}
& G_{x}=\int Dp\int_{x_{(in)}-\theta p/2\hbar}^{x_{\left( out\right)
}-\theta p/2\hbar}Dq\exp\left\{ \frac{i}{\hbar}S_{\mathrm{nonrel}}^{\theta
}\right\} \,,  \label{12} \\
& G_{p}=\overset{p_{out}}{\underset{p_{in}}{\int}}Dp\int Dq\exp\left\{ \frac{%
i}{\hbar}\tilde{S}_{\mathrm{nonrel}}^{\theta}\right\} \,,  \label{14}
\end{align}
where%
\begin{align}
S_{\mathrm{nonrel}}^{\theta} & =\int dt\left[ p_{j}\dot{q}^{j}-H\left(
p,q\right) +\dot{p}_{j}\theta^{ji}p_{i}/2\hbar\right] ,  \label{11a} \\
\tilde{S}_{\mathrm{nonrel}}^{\theta} & =\int dt\left[ -q_{j}\dot{p}%
^{j}-H\left( p,q\right) -p_{j}\theta^{ji}\dot{p}_{i}/2\hbar\right] .
\label{12a}
\end{align}

One ought to stress that the actions $S_{\mathrm{nonrel}}^{\theta}$ and $%
\tilde{S}_{\mathrm{nonrel}}^{\theta}$ differ by a total time derivative.

The path-integral (\ref{12}) is a generalization of the result obtained in 
\cite{Acatrinei} for arbitrary nonrelativistic system and without any
restrictions on the matrix $\mathbf{\theta}$. One ought to say that path
integrals on noncommutative plane for matrix elements of the evolution
operator in coherent state representations were studied in \cite{Smailagic}
and \cite{Tan}. They have specific forms which is difficult to compare with
our results.

In the convetional ''commutative'' nonsingular QM the action $S_{\mathrm{%
nonrel}}^{\theta}$ (at $\theta=0)$ is just the Hamiltonian action of the
classical system under consideration. The canonical quantization of this
action reproduces the initial QM of the system. In the noncommutative case
this action is modified by a new term $\dot{p}_{k}\theta^{kj}p_{j}/2\hbar$.
One can treat the action (\ref{11a}) as a the $\theta$-modified Hamiltonian
action of the classical system under consideration (see the Introduction).
This interpretation can be justified by the canonical quantization of the
action, see \cite{Deriglazov}.

\end{document}